# A SEARCH FOR WATER VAPOR PLUMES ON EUROPA USING SOFIA

Sparks, W.B.[1,2], Richter, M.[3], deWitt, C.[4], Montiel, E.[3,4], Dello Russo, N.[5], Grunsfeld, J.M.[6], McGrath, M.A.[1], Weaver, H.[5] Hand, K.P.[7], Bergeron, E.[2], Reach, W.[4]

[1] SETI Institute, 189 N Bernardo Ave, Mountain View, CA 94043, USA

[2] Space Telescope Science Institute, 3700 San Martin Drive, Baltimore, MD 21218, USA

[3] Universiy of California, Davis, 1 Shields Ave, Davis, CA 95616, USA

[4] SOFIA Science Center, NASA Ames Research Center, Mail Stop N232-12, P.O. Box 1,Moffett Field, CA 94035, USA

[5] Johns Hopkins Applied Physics Laboratory, 11100 Johns Hopkins Rd., Laurel, MD 20723-6099, USA

[6] NASA Goddard Space Flight Center (Emeritus), Code 600, 8800 Greenbelt Road, Greenbelt, MD 20771, USA

[7] Jet Propulsion Laboratory, California Institute of Technology, 4800 Oak Grove Drive, Pasadena, CA 91109, USA

Short title: SOFIA OBSERVATIONS OF EUROPA

Corresponding author: wsparks@seti.org

ABSTRACT

We present mid-infrared SOFIA/EXES spectroscopy of Europa, seeking direct evidence of the presence of water vapor arising from plumes venting from the surface of Europa. We place quantitatively useful upper limits on the strength of water vibrational-rotational emission lines. Conversion to water mass limits is dependent on the rotational temperature of the vapor. For low rotational temperature, the limits lie below the inferred water mass from previous HST plume observations. For higher temperatures, the limits are comparable. We also present coordinated HST transit observations obtained close in time to the SOFIA observations. There is evidence for a feature close to the location of the previously seen feature north of the crater Pwyll in one of the HST images, although it was not observable by EXES given its location. We conclude that if a water plume had been active at the time of the SOFIA observation, with the strength implied by previous HST observations, then under the right Earth atmospheric and geometric conditions, the plume could have been detected by EXES, however no IR water vibrational-rotational emission was detected.

KEYWORDS: Europa, cryovolcanism

1. INTRODUCTION

With good evidence for a deep, ice covered, saline ocean, Europa is one of the most compelling astrobiological targets in the Solar System (Hand et al. 2009). The induced magnetic field and youthful surface morphology point towards a present-day, saline global ocean of ~100 km depth (Anderson et al. 1998; Pappalardo et al. 1998; Kivelson et al. 2000; Zimmer et al. 2000; Hand & Chyba 2007; Hand et al 2007). With the needed chemical and energetic ingredients, Europa is one of the most plausible locations to seek extant life beyond the Earth (Hand et al. 2007; 2009). The means by which water is transported from the deep ocean to the frozen surface, if indeed it is, are unknown. Internal ice mechanics and localized heating may cause regions of melt water within the ice, which could modify the integrity of the ice and result in outflow of liquid water onto the surface. Alternatively, fractures within the ice may penetrate directly or indirectly to the liquid ocean, offering pathways for water that was once in the deep ocean to be deposited onto the surface. The timescales for such transport mechanisms are different, and may be either short for direct penetration through the ice, or relatively long if ice convection and localized chaos formation are required (Schmidt et al. 2011).

Empirically, if water can be observed flowing onto the Europa surface or outgassing from that surface in the form of plumes, we may advance our knowledge of the processes within the ocean and its ice crust, and seek to understand the global systematics that govern Europa's surface and ocean. Equally, with relatively straightforward access to recently liquid water, we would have an inviting target in the plumes for in-depth studies relating to dynamics, chemistry and composition, including any signs of biological processes, circumventing the immediate need to drill through many kilometers of ice.

Evidence for plumes was found using two different observing techniques with HST. Roth et al. (2014) found line emission from the dissociation products of water. Sparks et al. (2016) found evidence for off-limb FUV continuum absorption as Europa transited the smooth face of Jupiter. The evidence improved with the discovery of a repeating plume candidate at the same location (to within the uncertainties) as a thermal anomaly previously seen on the Europa night-side by the Galileo spacecraft (Spencer et al. 1999; Sparks et al. 2017). Independently, Jia et al. (2018) found in situ evidence of a plume on Europa at an altitude of ~400 km from a reanalysis of the Galileo magnetic and plasma wave data. The Jia et al. plume is located about 1000 km northwest of the Sparks et al. (2017) repeat plume candidate. McGrath & Sparks (2018) showed that the strongest ionospheric detection in Galileo radio occultation data on the E6a flyby ingress occultation is close to the repeat plume location; however, the E6b egress occultation close to the Jia plume location did not show an ionospheric detection. Trumbo et al. (2017; 2018) showed that the repeat-plume region has high thermal inertia relative to its surroundings, indicative of a surface with distinct characteristics that could be a signature of either internal activity or external deposition.

Several alternative mechanisms have been proposed to create plumes, including explosive release of dissolved volatile compounds as confining pressure is removed, or pressurization and fracturing due to expansion of ice as an enclosed water cavity freezes (Pappalardo et al.

1999; Schmidt et al. 2011; Walker & Schmidt 2015). Subsequent collapse of such a cavity may be implicated in chaos formation and surface outflow (Schmidt et al. 2011).

While the two initial independent HST observational approaches offer statistically significant evidence for the presence of plumes, they are nevertheless close to the limits of detectability by HST. Not only that, in the approach of Roth et al. (2014), the detection is of hydrogen and oxygen emission, while Sparks et al. (2016) find an unspecified continuum absorber, presumed to be water but not uniquely identified as such. The plasma wave and radio occultation evidence for a plume is also indirect, suggesting environmental disturbances consistent with the presence of a plume. However, with mid-infrared high-resolution spectroscopy of the second fundamental vibrational mode series around 6.3 μm, we have the potential to unambiguously and uniquely identify the presence of water vapor, if present, at Europa. Here we report on high-resolution mid-infrared spectroscopic observations obtained using the Stratospheric Observatory for Infrared Astronomy (SOFIA), which flies high enough above the terrestrial water vapor to enable detection of water vapor and other infrared emission from astronomical sources. We also present new Hubble Space Telescope (HST) transit observations that were acquired close in time to the SOFIA flights, to help tie the two approaches together. The SOFIA observations provide quantitatively comparable upper limits to the HST observations, and are shown to have the capacity to directly detect plumes on Europa at the level previously inferred, if active at the appropriate time.

2. OBSERVATIONS AND DATA PROCESSING

SOFIA carries a 2.7-m primary telescope to altitudes between 11.9 km (~39,000 ft) and 13.7 km (45,000 ft) to perform astronomical observations (Young et al. 2012). A principal investigator instrument, the Echelon Cross Echelle Spectrograph (EXES; Richter et al. 2010), is capable of observing between 4.8 and 28.3 μm. It is equipped with a 1024 x 1024 Si:As detector array, of similar design to the Texas Echelon Cross Echelle Spectrograph (TEXES; Lacy et al. 2002). EXES is able to perform high-resolution (up to R = 100,000) spectral observations due to a steeply blazed aluminum reflection grating used as an echelon in association with an echelle grating to cross-disperse the spectrum.

Observations of Europa were carried out during two different flights: 15 March 2017 and 27 May 2017 (see Table 1) as Guest Observer program 05_0153. The leading hemisphere of Europa was observed during the first flight, while the trailing hemisphere was observed during the second flight. The dates were chosen to provide sufficient Doppler shift of Europa relative to Earth to separate the target emission lines from terrestrial absorption lines, with the additional constraint of being able to observe with SOFIA near Europa's maximum elongation from Jupiter for a continuous period >3 hrs. The wavelength range covered was 6.101 – 6.146 μm (1627 – 1639 cm$^{-1}$) in the "High-Medium" configuration. Spectra of Polaris were also acquired to serve as telluric and flux calibrators. As an F star, Polaris does not have H$_2$O in its photospheric spectrum. The spatial PSF of SOFIA has a core with FWHM ~2.5" and we used a 2.4" wide slit. The angular diameter of Europa is less than 1", hence the satellite is a point source from EXES perspective. An unresolved line with the slit has a FWHM of 4.5 km/s.

The continuum of Europa in this spectral region is too weak for EXES/SOFIA to detect, preventing peak-up at this wavelength. However, by changing the order sorting filter, without moving any other optical elements, we were able to peak-up on Europa's 19 μm continuum.

This was done with the echelle angle fixed so that there were no adjustments from the desired science range. Peak-ups were done about once every hour in order to ensure that Europa remained well-centered on the spectrograph slit.

The data, Fig. 1, were processed using tools from the EXES Redux package (Clarke et al. 2015), based on routines developed for TEXES (Lacy et al. 2002.). The pipeline was run as normal (nod subtracted including a local sky subtraction for each spectrum order, flat corrected, rectified geometrically and coadded), with the exception that all nod pairs together were used to estimate the threshold for removing temporal spikes from individual pixels. For the May 2017 run, the first nod pair of each individual observation was not used due to its elevated noise. Custom routines were used to extract 1D spectra, **outlined below**. The sky emission spectrum was used to generate a 2-D wavenumber map for the observations, where the HITRAN (Gordon et al. 2017) wavenumbers of known telluric water emission lines were used to calibrate the dispersion and wavenumber zeropoint. The wavenumber solution is accurate to 0.5 km/s.

In coordination with this SOFIA program we acquired two additional HST observations (HST GO#14930), with a third serendipitous observation from another program (HST GO#14891 visit 1) providing the closest observation in time to the SOFIA March observing run. Europa was observed as it transited the smooth face of Jupiter using the far ultraviolet (FUV) multi-anode microchannel array (MAMA) in STIS in time-tag imaging mode, following Sparks et al. (2016). The F25SRF2 filter excludes geocoronal Lyman α and provides an effective wavelength of ~150 nm. The time-tag imaging mode provides a position and time for every detected photon event, with a time resolution of 125 μs. The details of the data acquisition and HST data processing procedures are described in Sparks et al. (2016). The images are shown in Fig. 2, following division by a model of the Jovian background light (Sparks et al. 2016). To quantify these images, we divided each of the three observations by the average of the other seventeen transits available to us at that time and converted to a standard normal deviate "*z*" image by subtracting the mean and dividing by the standard deviation.

4. RESULTS

For each of the two SOFIA/EXES observing runs, we created a single overall mean and uncertainty image for the two-dimensional spectroscopic data, following the basic reductions described above. We also modeled the Gaussian spatial profile *S(x)* of a point source, where *x* is the coordinate across the spectrum, at each location in the same two-dimensional data, using the known position of Europa. We then extracted spectra for individual EXES orders in two ways. A total flux $f_{tot}$ was evaluated at each wavenumber for points where the spatial response *S(x)* exceeds 0.2× its peak, with a small correction for flux excluded by the cutoff. We also derived a flux conserving optimally weighted flux at each wavelength (Horne 1986), for each order, where $f_{opt} = \sum f(x_i) w_{i1} / \sum w_{i2}$, where the *f(x$_i$)* are the flux values at position *x$_i$* across the spectrum , $w_{i1} = S(x_i)/\sigma_i^2$ and $w_{i2} = S^2(x_i)/\sigma_i^2$, where $\sigma_i$ is the uncertainty at each point. To improve stability, for each spatial profile we assumed the $\sigma_i$ was constant, equal to the mean uncertainty estimated from the stack of nod pairs for that profile. Spectra of Polaris obtained with the same configuration were used to correct for the atmospheric transmission and provide an absolute flux calibration.

Tables 2 and 3 list the parameters for the three strongest lines expected in the chosen wavelength ranges, and the results for flux measurements and mass estimates, for the leading and trailing hemispheres respectively. To provide a control region of the spectrum, we reversed the Doppler shift and measured the flux in regions of the spectrum where no lines are expected, and include these in the tables. The uncertainties tabulated are empirical measurements of the standard error of the pixel mean derived from the stack of nod pairs, averaged across the spatial profile used in the optimal extractions. A comparison of the empirical standard deviations in regions of the two-dimensional data to the uncertainties derived for the individual pixels from the nod-pair stack showed the former to be higher by 18.6% and 12.1% for the March and May 2017 runs, respectively. This is likely due to residual spatial structure in the two dimensional data arising (e.g.) from time variability in the terrestrial spectrum, detector features such as pattern noise, multiple amplifier readouts, and flat fielding defects. We included these correction terms in the noise estimates presented.

The resulting flux estimates to their uncertainties are in the range -2.3 to +2.7 $\sigma$, which we interpret as non-detections. While a 2.7 $\sigma$ result is nominally intriguing, it is not sufficient, in our opinion, to justify a claimed detection. The sky transparency is time variable and residual systematics as mentioned above demand a higher degree of certainty than afforded by such a measurement.

The $H_2O$ production rate released from Europa (Q, molecules s$^{-1}$) can be determined using the idealized assumption of uniform spherical outflow from the source (although in the optical thin case the geometry is not important – we are simply counting molecules):

$$Q = \frac{4\pi\Delta^2}{hc\nu(\tau \sum g_{line})_{1AU} f(x)} \sum F_{line}$$

where $\Delta$ (m) is the SOFIA-Europa distance, hcν is the energy per photon (J), $\tau$ and $g_{line}$ are the $H_2O$ lifetime (seconds) within the field of view and fluorescence g-factor (photon s$^{-1}$ molecule$^{-1}$) respectively at a heliocentric distance of 1 AU (Yamamoto 1982). The quantity $f(x)$ represents the fraction of the total number of released $H_2O$ gas molecules within the aperture. The upper limit of the flux from the summed $H_2O$ lines for March 2017 is $\Sigma F_{line}$ = 4.7 x 10$^{-15}$ erg s$^{-1}$ m$^{-2}$. The lifetime of $H_2O$ molecules within the SOFIA field of view is potentially governed by two processes: the photodissociation $H_2O$ in the solar radiation field, and the ballistic trajectory for a molecule of $H_2O$ to be ejected from Europa and fall back to its surface. The photodissociation lifetime of a $H_2O$ in the solar radiation field at 1 AU is ~ 7.7 x 10$^4$ s (Huebner et al. 1992), and at the heliocentric distance of Europa, the lifetime is about thirty times longer or ~ 2.3 x 10$^6$ s. The $H_2O$ lifetime in Europa's "atmosphere" is therefore expected to be controlled by the ballistic trajectory lifetime, which is much shorter (~ 1 x 10$^3$ s if the height of ejection is between ~ 100 – 200 km as required by the HST observations of plume candidates, Roth et al. 2014; Sparks et al. 2016). Hence we adopt the ballistic lifetime of 1000 s in our calculations of abundance.

Upper limits for line fluxes were determined for the three strongest $H_2O$ $\nu_2$ lines covered by SOFIA: $2_{02}$-$1_{01}$ (1627.827 cm$^{-1}$), $1_{11}$-$0_{00}$ (1634.967 cm$^{-1}$), and $3_{12}$-$3_{03}$ (1635.652 cm$^{-1}$). For reasonable values of the gas rotational temperature $T_{rot}$, the combined g-factors for these lines range from ~ 5.5 x 10$^{-5}$ (10 K) – 8.6 x 10$^{-6}$ (100 K) photon s$^{-1}$ molecule$^{-1}$ (Villanueva et al. 2012), and spectral outputs at different temperatures from the NASA-GSFC planetary spectrum generator (Villanueva et al. 2018). The spatial PSF of SOFIA is much larger than size of Europa during these observations; therefore, we assume that all released $H_2O$ molecules were within the SOFIA field of view, $f(x) = 1$. Based on these parameters and assumptions we derive an upper limit of Q < 1 x 10$^{30}$ molecules s$^{-1}$ assuming a gas rotational temperature = 100 K, and Q < 2.5 x 10$^{28}$ molecules s$^{-1}$ (Q < 750 kg s$^{-1}$) for a gas rotational temperature $T_{rot}$ = 10 K, that is a total number of molecules (independent of timescale) $N_{mol}$ < 1 x 10$^{33}$ for a rotational temperature $T_{rot}$ = 100 K, and $N_{mol}$ < 2.5 x $10^{31}$ for $T_{rot}$ = 10 K.

The HST $z$-images were inspected for outliers relative to an empirical average of many images, which would indicate the presence of plume candidates as described above. With this simplified approach, the image obtained on March 12, 2017 shows two compact, marginally significant features, at $z$≈3.4 latitude -18° on the trailing hemisphere, and another almost diametrically opposite on the leading hemisphere. The location of the trailing hemisphere candidate is less than 2 pixels from the previously identified candidates near the crater Pwyll (Sparks et al. 2017), consistent within the uncertainties, and adding to statistical evidence for a plume at this position. The implied optical depths are τ≈0.2, similar to the values found in Sparks et al. (2016), where the March 17, 2014 candidate had a peak optical depth of 0.34 and covered approximately twice the area. The estimate for the number of $H_2O$ molecules for that event was ≈2 x $10^{32}$ molecules, hence at face value, the new observation would require 4 x $10^{31}$ molecules. The other two images showed no absorption patches with $z$ values greater than 3, suggesting by comparison to the previous results, less than or of order $10^{31}$ molecules in limb plume candidates at those times. There is a similarly marginal positive feature at 33°S, 185 km from the limb in the third of the image sequence, which, we speculate, may be related to auroral activity ([OI] emission lines are within the bandpass), see Roth et al. (2016). The SOFIA/EXES and HST analyses are encouragingly close to one another, with consistent estimates and limits. The HST observations will be consolidated with additional ongoing observations, where a rigorous statistical study, beyond the scope of the current letter, will be provided.

## 5. DISCUSSION AND CONCLUSIONS

The features identified in the work of Roth et al. (2014) and Sparks et al. (2016) both required a total of approximately 1–2×$10^{32}$ $H_2O$ molecules, 3–6×$10^6$ kg, which if active for a 1000 sec ballistic timescale is ~3–6×$10^3$ kg s$^{-1}$. To grasp this intuitively, one standard Olympic swimming pool has a mass of 2.5×$10^6$ kg of water. The mass constraints provided by the SOFIA EXES measurements depend significantly on the rotational temperature of the putative water vapor. For Europa this is unknown; however, measurements are available for many comets, with temperature primarily dependent on the gas production rate and typically in the range $T_{rot}$ ~20–150K. Most rotational temperature measurements of comets are obtained when the comet is within 2 AU of the Sun. The few comets measured at larger heliocentric distance show lower rotational temperatures. Comet 29P/Schwassmann-Wachmann at a heliocentric distance of 6.2 AU was at a comparable distance from the Sun as Europa, and had a loss rate of Q~2×$10^{28}$ molecules s$^{-1}$ and measured rotational temperature of only $T_{rot}$≈5K, based on CO (Gunnarsson et al. 2008; Paganini et al. 2013). Rotational temperatures measured for CO, $CH_4$, and $C_2H_6$ in comet C/2006 W3 Christensen were between ~ 10 – 25 K at $R_h$ ~ 3.25 AU for Q~3×$10^{28}$ molecules s$^{-1}$ (Bonev et al. 2017). Hence the assumed range for Europa is reasonable, with the more likely value being closer to $T_{rot}$ ≈ 10 K based on a cometary analogy.

The HST images taken around the time of the SOFIA EXES spectra, Fig. 2, show possible features in the March 12 image, one close to the repeating plume candidate of Sparks et al. (2017). Since this appears at the trailing edge of Europa (longitude ≈270°), and in March EXES observed the leading hemisphere (sub-longitude 92°), EXES would not have been able to observe this plume candidate, were it also active at the time of the EXES observation. The other potential candidate diametrically opposite is statistically marginal but would have been within the EXES field of view at a level close to the limits of the EXES data. The HST image closest in time to the trailing hemisphere observation does not show any evidence of plumes, nor does the final image, taken some significant time after the SOFIA run. With an estimated plume duty cycle of 17% (Sparks et al. 2017), and ballistic timescales of only 1000 sec, these statistics are consistent with the known intermittency of the phenomenon.

If plumes can be securely identified on Europa, and quantitatively characterized, the implications will be profound. Direct infrared spectroscopic observations may have a very important role to play: EXES on SOFIA has the potential to unambiguously do so at a level consistent with the inferred mass estimates of the prior HST observations. With plume activity, we can probe water that has recently been in the liquid state, and seek to determine where it originated and ultimately, what other compounds it contains, and hence we may advance our understanding of the implications for biology at Europa.

ACKNOWLEDGEMENTS

This work is based on observations made with the NASA/DLR Stratospheric Observatory for Infrared Astronomy (SOFIA). SOFIA is jointly operated by the Universities Space Research Association, Inc. (USRA), under NASA contract NNA17BF53C, and the Deutsches SOFIA Institut (DSI) under DLR contract 50 OK 0901 to the University of Stuttgart. Financial support for this work was provided by NASA through award SOF 05_0153 issued by USRA. The FUV imaging data were obtained using the Hubble Space Telescope which is operated by STScI/AURA under grant NAS5-26555. We acknowledge support from a grant associated with HST award GO-14930. KPH acknowledges support from the Jet Propulsion Laboratory, California Institute of Technology, under a contract with the National Aeronautics and Space Administration.

TABLE 1 OBSERVING LOG

|  | Date | UT range | Exposure time (s)* | Altitude range (feet) | Program |
|---|---|---|---|---|---|
| SOFIA/EXES | 03/15/2017 | 07:01–10:19 | 8640 | 42000–45000 |  |
|  | 05/27/2017 | 04:04–07:02 | 7936 | 39000–42000 |  |
|  |  |  |  |  |  |
| HST/STIS/F25SRF2 | 03/12/2017 | 15:37–16:39 | 2650 |  | 14891 |
|  | 05/26/2017 | 03:49–04:50 | 1778 |  | 14930 |
|  | 06/12/2017 | 21:50–22:51 | 1403 |  | 14930 |

*On source time, total time is twice this, half on source, half on sky, for EXES.

TABLE 2 LEADING HEMIPSHERE LINES AND FLUX LIMITS – MARCH 2017

| Rest wavenumber ($cm^{-1}$) | Wavenumber range ($cm^{-1}$) | Flux (erg $s^{-1}cm^{-2}$) | Flux uncertainty (erg $s^{-1}cm^{-2}$) | Flux/uncertainty N(sigma) | Molecules implied ($T_{rot}$=10–100K) |
|---|---|---|---|---|---|
| Europa | v≈-26.45 km $s^{-1}$ |  |  |  |  |
| 1627.8274 | 1627.9602 - 1627.981 | 4.70e-015 | 1.72e-015 | 2.74 | 3e32 – 6e32 |
| 1634.9671 | 1635.1004 - 1635.1222 | -4.01e-015 | 2.25e-015 | -1.78 | 3e31 – 4e32 |
| 1635.6518 | 1635.7852 - 1635.8071 | -8.66e-015 | 3.78e-015 | -2.29 | 7e32 – 2e34 |
| Control |  |  |  |  |  |
| 1627.8274 | 1627.6729 - 1627.6947 | -1.30e-015 | 2.37e-015 | -0.55 |  |
| 1634.9671 | 1634.8119 - 1634.8337 | 3.04e-015 | 2.24e-015 | 1.36 |  |
| 1635.6518 | 1635.4966 - 1635.5184 | -1.22e-014 | 8.49e-015 | -1.44 |  |

TABLE 3 TRAILING HEMISPHERE LINES AND FLUX LIMITS – MAY 2017

| Rest wavenumber ($cm^{-1}$) | Wavenumber range ($cm^{-1}$) | Flux (erg $s^{-1}cm^{-2}$) | Flux uncertainty (erg $s^{-1}cm^{-2}$) | Flux/uncertainty N(sigma) | Molecules implied ($T_{rot}$=10–100K) |
|---|---|---|---|---|---|
| Europa | v≈+34.4 km $s^{-1}$ |  |  |  |  |
| 1627.8274 | 1627.6298 - 1627.6515 | -5.44e-015 | 2.73e-015 | -1.99 | 5e32 – 9e32 |
| 1634.9671 | 1634.7686 - 1634.7904 | 2.13e-015 | 3.42e-015 | 0.62 | 4e31 – 6e32 |
| 1635.6518 | 1635.4532 - 1635.4751 | 1.63e-015 | 2.79e-015 | 0.58 | 5e32 – 2e34 |
| Control |  |  |  |  |  |
| 1627.8274 | 1628.0034 - 1628.0251 | 4.17e-015 | 2.84e-015 | 1.47 |  |
| 1634.9671 | 1635.1438 - 1635.1656 | -2.42e-015 | 3.86e-015 | -0.63 |  |
| 1635.6518 | 1635.8286 - 1635.8504 | -7.70e-017 | 5.42e-015 | -0.01 |  |

FIGURES:

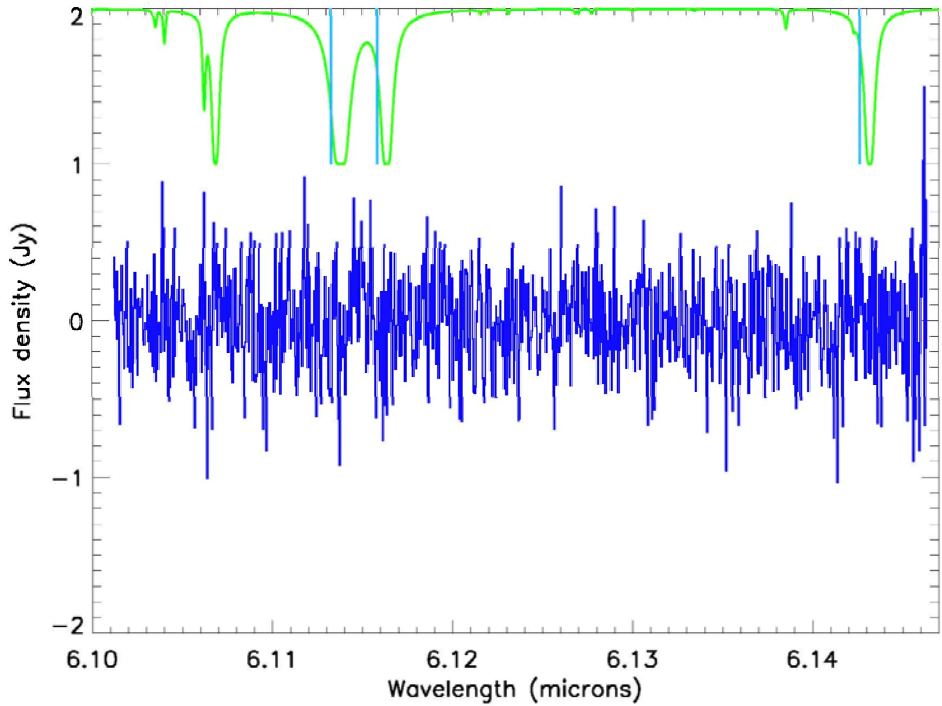

Fig. 1. EXES spectrum showing Nyquist sampled binned spectrum with the 4.5kms$^{-1}$ slit for March 2017. The ATRAN model (green) is matched to the transmission in the adjacent Polaris leg at the same altitude. (4μm overburden) and is offset by +1 Jy for clarity (i.e. the transmission is 0% at y=1 and 100% at y=2)) with Doppler shifted line positions (blue) shown.

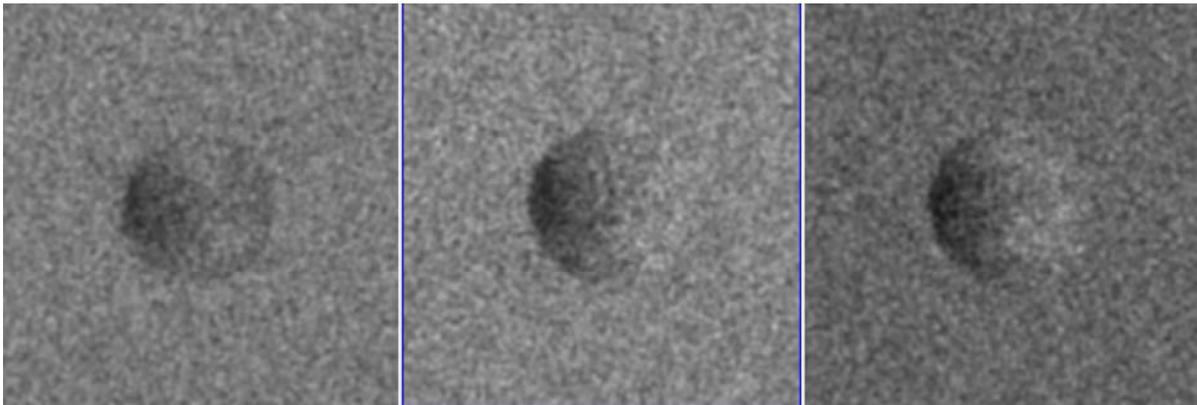

Fig. 2. HST transit images nearest in time to the SOFIA EXES observations. From left to right, 3/12/17 (2.7 days), 5/26/17 (1.0 days), 6/12/17 (16.7 days). HST images have Europa-north up, trailing hemisphere to the left, against the smooth background of Jupiter. The leftmost image is closest to the leading hemisphere SOFIA observation, while the central image is closest to the trailing hemisphere SOFIA observation.